\documentclass{caps}
\usepackage{peter}
\usepackage{epsf}
\begin{document}
\pagenumbering{arabic}

\def\apj{\textit{ApJ}}
\def\apjl{\textit{ApJ Lett.}}
\def\apjs{\textit{ApJ. Supp.}}
\def\aap{\textit{Astronomy \& Astrophysics}}
\def\aj{\textit{AJ}}
\def\nat{\textit{Nature}}
\def\mnras{\textit{MNRAS}}
\def \lta {\mathrel{\vcenter
     {\hbox{$<$}\nointerlineskip\hbox{$\sim$}}}} 
\def\gcc{{\rm g cm$^{-3}$}}
\def\gcm3{g~cm$^{-3}$}
\def\g-s{g~s$^{-1}$}
\def\cm3s{cm$^3$~s$^{-1}$}
\def\cms{cm~s$^{-1}$}
\def\kms{km~s$^{-1}$}
\def\erg-s{erg~s$^{-1}$}
\def\degree{$^{\rm o}$}
\def\beq{\begin{equation}}
\def\eeq{\end{equation}}
\def\gr{$\gamma$-ray}
\def\grs{$\gamma$-rays}
\def\grb{$\gamma$-ray burst}
\def\grbs{$\gamma$-ray bursts}


\author{J. Craig Wheeler\\Department of Astronomy\\
University of Texas at Austin
}

\chapter{3-D Explosions:\\ A Meditation on Rotation (and Magnetic Fields)
}
\section{Introduction: A Brief Time for History}

There has been a great deal of progress in the thirty-five years
or so that I have been working on supernovae and related topics.  Two
of the classical problems have been with us the whole time: what
makes core collapse explode, and what are the progenitors of
Type Ia supernovae?  This workshop, indeed, the perspectives of
three-dimensional astrophysics applied to these problems, gave
encouraging evidence that breakthroughs may be made in both of
these venerable areas.

On the other hand, what a marvelous array of progress has rolled
forth with ever increasing speed.  We have an expanded botany of
supernovae classification: Type Ia, Ib, Ic,Type IIP, IIL IIb,
IIn; but, of course, more than mere classification, a growing
understanding of the physical implications of these categories.
Neutron stars were discovered as rotating, magnetized pulsars
when I was a graduate student, and the extreme form, magnetars,
has now been revealed (Duncan \& Thompson 1992).
The evidence that we are seeing black
holes in binary systems and the centers of  galaxies has grown
from suspicion to virtual certainty, awaiting only the final
nail of detecting the black spot in a swirl of high-gravity
effects.  Supernova 1987A erupted upon us over 16 years ago and
is still teaching us important lessons as it reveals its
distorted ejecta and converts to a young supernova remnant
before our eyes.  

There have also been immense theoretical developments.  Focus on
core collapse has stimulated so much great work on neutrino
transport: the invocation of weak neutral currents and
neutrino-nucleon scattering; the understanding that neutrinos
can and will become degenerate at the highest densities and the
concomitant implications for the dynamics and the formation of the
homologous core.  More recently we have come to general
understanding that a prompt shock is unlikely to make an
explosion, but that significant layers of the proto-neutron star
will be convective with important implications for the neutrino
transport.  Techniques of neutrino transport have evolved from
simple diffusion to full Boltzmann transport.  SN 1987A showed
dramatically that we are on the right track, even if the
details, even important physics, may be missing: core collapse
with the predicted production of neutrinos does occur!   In
terms of the ``other mechanism," our understanding has evolved
from detonations to deflagrations, to the current paradigm of
delayed detonation models.  The recent understanding of the
associated combustion physics has blossomed with the
computational ability to do the required three-dimensional
modeling.  

Finally, the last few years have seen the birth and maturation
of a field that was hinted at long ago, but came to fruition
only recently, the systematic study of the polarization of
supernovae.  This technique has substantially altered our view
of core collapse.  It was only a few years ago that polarization
was still regarded as an oddity, perhaps limited to a few
peculiar events.  In the last year, the idea that core collapse
is asymmetric has become sufficiently accepted that papers are
now written saying ``as is well known, core collapse is
asymmetric" without providing any reference to the hard labor
required to establish that!  Overnight, it seems, the wonders of
the three-dimensional world have become revealed wisdom.  The
revelations of polarized core collapse have been the most
distinct so far, but their implications are far from
understood.  The application of polarization to Type Ia
supernovae had lagged somewhat in drama because the polarization
is generally smaller, but this workshop served to provide
evidence that important three-dimensional distortions are
ubiquitous, and important, in Type Ia as well.

Besides all these developments that have been so central to the
development of supernova science, the last few years have seen
two outstanding developments that have cast supernovae research,
already one of the most central and important in astrophysics,
onto broader stages.  What a time was 1997/1998!  Careful studies
of Type Ia supernovae revealed the acceleration of the Universe with
the implication of the pervading dark energy.  In virtually
the same time frame, the discovery of optical transients associated
with gamma-ray bursts and then SN 1998bw led to the connection of
gamma-ray bursts with supernovae, probably some variety of Type Ic.
For a mature field, the study of supernovae had a great deal of life
left!  Since Type Ia and Type Ic have been especially near and dear to
me, this was about more excitement than my mature heart could
stand.  

I cannot  do justice to all the great work on supernovae that has 
been done over my career, but I would like to touch on one other bit 
of history, a development that was critical for so much else that 
followed.  I distinctly recall that when I was in graduate school 
there was a raging debate concerning the nature of the spectra of
Type Ia, then called just Type I, supernovae.  Some people
argued that the spectrum near maximum consisted only of
absorption lines and provided the interpretation of the
absorption minima in terms of atomic features.  Others insisted
that the spectrum consisted purely of emission lines and
provided an interpretation of the flux peaks, totally
incongruent with the first interpretation, of course.  David
Branch provided the insight that we were  looking at P-Cygni
lines, hence a blended mix of emission and blue-shifted
absorption.  That was the insight needed to convince the world
that the key feature in the spectrum of a Type Ia was Si II.  From 
that it followed that the presence of silicon and other
intermediate mass elements ruled out pure detonation models.
This was the base on which so much subsequent analysis of
supernovae of all types was built.   More work, especially from
Bob Kirshner and colleagues revealed that, with
patience, the spectrum does evolve to be dominated by emission
lines.  Type Ia, like all supernovae, eventually evolve to a
``supernebular" phase.

\section{Type Ia}

The combination of ever more thorough searches both by people at
the eyepiece and by computer-driven telescopes, subsequent
multi-wavelength follow-up, and theoretical and computational
study  has brought the study of Type Ia supernova to an
impressive level of maturity.  After a spirited debate, the
conclusion  that Type Ia are not merely thermonuclear explosions
in white dwarfs, but specifically explosions in carbon/oxygen
white dwarfs of mass very nearly the Chandrasekhar mass is now
essentially universally accepted (H\"oflich \& Khokhlov 1996;
Nugent et al. 1997; Lentz et al. 2001).  Even more precisely,
the paradigm of a slow initial subsonic deflagration phase followed
by a rapid supersonic, shock-mediated detonation phase (Khokhlov 1991)
has been richly successful in accounting for the observed properties
of Type Ia (H\"oflich 1995).  It accounts for the existence of
iron-peak elements in the center of the explosion and layers of
intermediate Ðmass elements in the outer layers, essentially by
design.  It also gives a framework in which to understand the
variety of light curve shapes with lower transition densities
leading to less nickel, and dimmer, cooler, faster light curves
(H\"oflich et al. 1996), and it has successfully made predictions
about infrared spectra (H\"oflich et al. 2002) and polarization
properties (Wang, Wheeler \& H\"oflich 1997; Howell et al. 2001).  
{\it Delayed detonation works!}

This success has put focus on a wonderful physics problem, the
deflagration to detonation transition, or DDT, that astrophysics
shares with a host of terrestrial combustion issues.  This is a
hard problem on Earth or off!  One of the most interesting
developments in recent years has been the resonance of
terrestrial and astrophysical combustion studies.  There has
been dramatic progress in understanding DDT in laboratory,
shock-tube environments by means of sophisticated computational
studies of shock-flame interactions (Khokhlov \& Oran 1999) and
DDT in enclosed environments where boundaries and reflected shocks
play a key role (Khokhlov, Oran \& Thomas 1999). Still, the
astrophysical problem, one of unconfined DDT, remains  elusive.
This is a quintessential multi-dimensional problem, one for
which several promising lines of attack are underway.

The wealth of knowledge of Type Ia revealed by optical studies
is too large to summarize here, but it has been amplified and
complemented in recent years by studies in the near infra-red.
The NIR is an especially powerful spectral range to study
because lines are less blended and the continuum is nearly
transparent so one sees all the way through the ejecta with a
single spectrum probing all the important layers
simultaneously.  This technique was pioneered for all supernovae
by Peter Miekle and his collaborators and is rapidly coming to
the fore as a major tool in the study of Type Ia.  SN 1999by was
a subluminous Type Ia that was, not incidentally, significantly
polarized (Howell et al. 2001). H\"oflich et al. (2002) showed
that a delayed detonation model selected to match the light
curve provided a good agreement with the NIR spectra and
revealed the products of explosive carbon burning in the outer
layers and products of incomplete silicon burning in deeper
layers.  The results were inconsistent with pure deflagration
models or merger models that leave substantial unburned matter
on the outside.  The data also seemed incompatible with the
mixing of unburned elements into the center as predicted by pure
deflagration 3-D models.  Three-dimensional models in which the
inner unburned matter undergoes a detonation, the current most
realistic manifestation of the delayed detonation paradigm as
presented here by Gamezo et al. alleviate that problem.
Marion et al.  (2003) have presented NIR spectra of ``normal"
Type Ia (see also Hamuy 2002) and shown that the outer layers of
intermediate mass elements are not mixed, that very little unburned
carbon remains in the outer layers, and perhaps revealed Mn,
a sensitive probe of burning conditions.

Another important development concerns work on the
quasi-static phase of carbon burning that follows carbon
ignition and precedes dynamic runaway.  This important
``smoldering" phase had not been critically re-examined since the
initial study of Arnett (1969).  H\"oflich \& Stein (2002) showed that
the convective velocities in this phase can exceed the initial
speeds of the subsequent deflagration front.  This means that
the ``pre-processing" of the white dwarf by this smoldering phase
and the resulting velocity field, rather than the pure
Rayleigh-Taylor driven deflagration, will dominate  the early
propagation of the burning front.  This is a crucial,
multidimensional, insight that will foment much work in the
near future to understand all the implications.

Finally, it is necessary to repeat that polarization studies
have revealed that Type Ia are polarized and hence asymmetric
(Wang, Wheeler \& H\"oflich 1997).  It may be that the subluminous 
variety are more highly polarized and perhaps more rapidlly rotating
than the ``normal" type (Howell et al. 2001).  It may also be
that, although the polarization is generally low, all Type Ia
are polarized at an interesting level if appropriate, sufficiently
accurate observations are made (Wang et al. 2003a).  This has clear
implications for the quest to answer the old problem of whether
Type Ia arise in binary systems and, if so, as we all believe,
what sort?  The asymmetries might also be teaching us lessons yet
ungleaned about the combustion process which is undoubtedly complex
and three dimensional.  The asymmetries must be understood in order
to use Type Ia with great confidence as we move to the next
phase of cosmological studies where exceptionally precise
photometry and tight control of systematic effects will be
necessary to probe the equation of state of the dark energy.

In any case, the lesson of recent history and of this workshop
is that Type Ia supernovae are three dimensional!

\section{Asymmetric Core Collapse}

If anything, the polarization studies have had even more
dramatic impact on core collapse supernovae.  All core collapse
supernovae adequately observed are found to be polarized and
hence asymmetric in some way (Wang et al. 1996; Wang et al. 2001,
2002, 2003b; Leonard et al. 2000; Leonard \& Filippenko 2001;
Leonard et al. 2001, 2002).  Many of these events are
substantially bi-polar (Wang et al. 2001).  The fact that the
polarization is higher as one sees deeper in and is higher when
the hydrogen envelope is less, strongly indicates  that the very
machine of the explosion deep in the stellar core is asymmetric
and probably predominantly bi-polar.  SN 1987A reveals similar
evidence (Wang et al. 2002).  Other famous ``spherically
symmetric" supernovae are those that gave rise to the Crab Nebula
and to Cas A. 

Complementary computational  work has shown that jet-induced
explosions can produce the qualitative asymmetries that are
observed (Khokhlov et al. 1999, see also MacFadyen et al. 2001
Zhang, et al. 2003).
Khokhlov \& H\"oflich (2001) and H\"oflich, Khokhlov \& Wang (2001)
have shown that
asymmetric nickel deposition by a jet-like flow can produce
polarization  by asymmetric heating and ionization even in an
otherwise spherically-symmetric density distribution.  This very
plausibly accounts for the early low polarization in Type II
supernovae that grows as the underlying asymmetry is revealed.

The large question remains as to what causes the jet-like flow.
My bet is that this involves rotation and magnetic fields at the
deepest  level.  Rotatation alone can affect neutrino
deposition, but the case can be made that rotation without
magnetic fields is highly unlikely.  Akiyama et al.
(2003) have presented a proof of principle that the physics of
the magneto-rotational  instability (MRI: Balbus \& Hawley 1991, 1998)
is inevitable in the context of the differentially-rotating
environment of proto-neutron stars. The magnetic fields can in
turn affect the neutrino transport.  The ultimate problem of
core collapse is intrinsically three-dimensional involving rotation,
magnetic fields, and neutrino transport.  We have known this all
along (despite, not because of, cheap shots after core collapse
talks in which some joker  always asks ``but what about
rotation?" or ``but what about magnetic fields?"), but the new
polarization observations demand a new, integrated view.  This
makes a devilishly hard problem even harder.  Progress will come
by isolating and understanding pieces of the problem and
eventually sticking them together.

\section{The Magneto-rotational Instability and Core Collapse}

The advantage of the MRI to generate magnetic field is that while it
works on the rotation time scale of $\Omega^{-1}$ (as does field-line
wrapping), the strength of the field grows exponentially.  This means that
from a plausible seed field of $10^{10}$ to $10^{12}$ G that might result
from field compression during collapse, only $\sim$ 7 - 12 e-folds are
necessary to grow to a field of $10^{15}$  G. That is only
$(7 - 12)/2\pi \sim  1 - 2$ full rotations or $\sim$ 10 - 20 ms for
expected initial rotation periods of order 10 ms.  Furthermore, while
the growth time may depend on the seed field, the final saturation field
is independent of the seed field (unlike a linear wrapping model
that ignores the complications of reconnection, see Wheeler et al. 2000,
2002, for examples and other references).

Core collapse will lead to strong differential rotation near the
surface of the proto-neutron star even for initial solid-body
rotation of the iron core (Kotake, Yamada \& Sato 2003; Ott et al. 2003).
The criterion for instability to the MRI is a negative gradient
in angular velocity, as opposed to a negative gradient in
angular momentum for the Rayleigh dynamical instability. This
condition is broadly satisfied at the surface of a newly formed
neutron star during core collapse and so the growth of magnetic
field by the action of the MRI is inevitable.  More quantitatively,
when the magnetic field is small and/or the wavelength is long
(k $\mathrm{v_a}$ $<$ $\Omega$) the instability condition can be written
(Balbus \& Hawley 1991, 1998):
\begin{equation}
N^2+\frac{\partial\Omega^2}{\partial \mbox{ ln }\mathrm r} < 0,
\end{equation}
where N is the Brunt-V\"{a}is\"{a}l\"{a} frequency.  Convective
stability will tend to stabilize the MRI, and convective
instability to reinforce the MRI.  The saturation field given by
general considerations and simulations is approximately given by
the condition: $\mathrm{v_a}\sim$ $\lambda \Omega$ where
$\lambda \lta$ r or $\mathrm{B^2}\sim 4\pi \rho\mathrm{r^2\Omega^2}$
where $\mathrm{v_a}$ is the Alfv\'en velocity.

These physical properties were illustrated in the
calculations of Akiyama et al. (2003) who used
a spherically-symmetric collapse code to compute the expected
conditions, instability, field growth and saturation.  Akiyama et al.
assumed initial rotation profiles, solid body or differential, invoked
conservation of angular momentum on shells that should, at least, give
some idea of conditions in the equatorial plane and computed regions
of MRI instability.  They assumed exponential growth to saturation.
For sub-Keplerian post-collapse  rotation, Akiyama et al. found
that fields can be expected to grow to $10^{15}$  to $10^{16}$
G in a few tens of milliseconds.  The resulting
characteristic MHD luminosity (cf. Blandford \& Payne 1982) is:
\begin{eqnarray}
\mathrm L_{\mathrm{MHD}}&\sim& \mathrm B^2 \mathrm r^3\Omega/2\sim 3\times
10^{52}\mbox{ erg } \mathrm s^{-1} \mathrm B_{16}^2 \mathrm
R_{\mathrm{NS,6}}^3 \left (\frac{\mathrm
P_{\mathrm{NS}}}{10~\mbox{ms}}\right
)^{-1} \\\nonumber
&\sim&10^{51}-10^{52}\mbox{ erg } \mathrm s^{-1}.
\end{eqnarray}
If this power can last for a significant fraction of a second, a
supernova could result.  Figure 1 shows the results 
for a model
in which the iron core began with a smoothly decreasing distribution
of angular velocity and a central value of $\Omega = 1$  s$^{-1}$.

\begin{figure}[htp]
\begin{center}
\leavevmode\epsfxsize=10cm \epsfbox{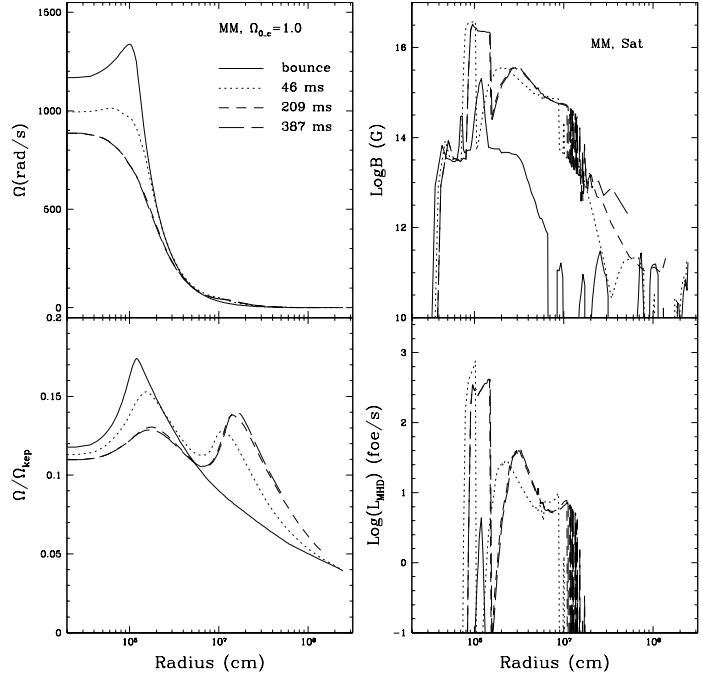}
\end{center}
\caption[mri.eps]{Angular velocity, field strength and MHD luminosity
(in units of $10^{51}$ erg s$^{-1}$) for a representative initial differential
rotation of the iron core as a function of time from Akiyama et al. (2003)}
\end{figure}



The implication of the work of Akiyama et al. (2003) is that
the MRI is unavoidable in the core collapse ambience, as
pertains to either supernovae or \grbs.  The field
generated by the MRI must be included in any self-consistent
calculation.  These implications need to be explored in much
greater depth, but there is at least some possibility that the
MRI may lead to strong MHD jets by the magneto-rotational
(Meier, Koide \& Uchida 2001) or other mechanisms.  A key point is that the
relevant dynamics will be dictated by large, predominantly toroidal
fields that are generated internally, not the product of twisting of
external field lines that is the basis for so much work on MHD
jet and wind mechanisms.  Understanding the role of these
internal toroidal fields in producing jets, in providing the
ultimate  dipole field strength for both ordinary pulsars and
magnetars (Duncan \& Thompson 1992),
in setting the ``initial" pulsar spin rate after the
supernova dissipates (that is, the ``final" spin rate from the
supernova dynamicists point of view), and any connection to
\grbs\ is in its infancy.

\section{Gamma-Ray Bursts}

    We have learned so much about gamma-ray  bursts since the
revolution of the discovery of afterglows it is impossible to do
it justice.  Briefly, we now know that the energy is not spread
isotropically, but is collimated into jets that  seem to have a
canonical energy of a few $\times 10^{50}$ ergs (Panaitescu \&
Kumar 2001; Frail et al. 2001). There is growing circumstantial
evidence for a connection to massive stars, yet there is
evidence that some gamma-ray  bursts explode  into a rather low
density ISM, and little evidence in many cases for the winds that should
characterize massive stars (Panaitescu \& Kumar 2002).

Now we have the dramatic evidence  from GRB030329/SN 2003dh
(Stanek et al. 2003; Hjorth et al. 2003;
Kawabata et al. 2003) of a definite connection between this burst and
a Type Ic-like supernova.  We also have the startling result of
the observation by Coburn \& Boggs (2003) of a large
polarization, $80\pm 20\%$ in GRB021206.  One interpretation of this
is that the Alfv\'{e}n speed considerably exceeds the sound
speed, implying a dynamically dominant field (Lyutikov,
Blandford \& Pariev 2003).

Despite these dramatic developments, there is much to be done.
In the context of gravitational collapse models we must consider
Keplerian shear, nearly equipartition fields, magnetic neutrino cross 
sections, strong magnetic helicity currents and viscoelastic effects
(see my concluding remarks), and a host of other effects that will 
pertain to this rapidly rotating and inevitably magnetic environment.

One of the issues currently facing the supernova community as we
grapple with the supernova/gamma-ray burst connection is whether
or not there is a new class of explosions (``hypernovae," see,
e.g. Maeda et al. 2002 and in these proceedings)  or if
the events we see with large photospheric velocities are just an
extension of a population with a continuum of properties.  There
are several complications  that must be borne in mind.  First,
the velocity at the photosphere is a very sensitive function of
time for Type Ic supernovae.  Just to say a velocity is ``high"
is not a terribly useful statement.  It is the whole velocity
evolution that must be compared to make a valid contrast of one
event with another.  Even then different velocity evolution can
and will result from different envelope masses without
substantial differences in explosion energy.  Another key factor
is that we now have substantial evidence, some of it quite
direct, that Type Ic supernovae are strongly asymmetric.  This
means that we might see different photospheric velocities in
different directions (H\"oflich, Wheeler \& Wang 1999).  
We can also see different luminosities in
different directions. This is then related to the deduction of
nickel masses, a key factor in the definition, at least in some
cases, of ``hypernovae."   In addition, non-spherical explosions
can affect the resulting density distribution and hence the
gamma-ray deposition and even the late time luminosity.  Since
explosion energies and nickel masses are quantities derived from
spherical models, great care must be taken in their
interpretation when strong asymmetries  are suspected.

To illustrate the empirical case, Table 1 gives photospheric velocity 
at maximum light and peak brightness for a sample of Type Ic and 
related supernovae for which such data were available.  Whether
or not a comparison at maximum  light is valid or the best way to do
this is  not clear.  The very fact that the data are sparse is a
cautionary note to both advocates and critics of ``hypernovae."
Nevertheless, this table illustrates that Type Ic come with a
considerable  dispersion in both peak brightness and
photospheric  velocity.  While it may be that SN 1998bw and SN
2003dh are especially bright, there are others as bright; and
while those events may have shown high velocities, so have other
Type Ic with modest peak brightness.  Further data may reveal
differently, but this table reveals no special pattern nor
obvious bifurcation in properties between normal Type Ic
supernovae and ``hypernovae."

\begin{table}

\begin{tabular}{|l|c|c|c|}
\hline
EVENT & Peak M$_V$ &  v at Peak & Ref \\
    &    & (1000 km~s$^{-1}$ ) &\\
\hline\hline

SN 1983V    &  18.1       &     15        &      (1)\\

SN 1983N    &   17.4       &     10      &        (2)\\

SN 1987K    &   16.9       &     10        &      (3)\\

SN 1987M      & 18.5       &     10        &      (4)\\

SN 1992ar  &    19.3        &    15        &      (5)\\

SN 1994I   &    18.1      &      14       &       (6, 7)\\

SN 1993J     &  17.7      &      10        &      (8)\\

SN 1997ef  &    17.2      &      11      &        (9)\\

SN 1998bw   &   19.4      &      15        &      (10, 11)\\

SN 2002ap   &   17.7        &    15        &      (12)\\

SN 2003dh   &   20.5        &    23       &       (13)\\

\hline
\end{tabular}

\noindent

{\footnotesize REFERENCES - (1) Clocchiatti et al. (1997);
(2) Clocchiatti, Wheeler, Benetti \& Frueh (1996);
(3) Filippenko (1988); (4) Filippenko, Porter \& Sargent (1990);
(5) Clocchiatti et al. (2000); (6) Richmond et al. (1996);
(7) Millard et al. (1999); (8) Wheeler \& Filippenko (1996);
(9) Mazzali, Iwamoto \& Nomoto (2000); (10) Galama et al. (1998);
(11) Patat et al. (2001); (12) Gal-Yam, Ofek \& Shemmer (2002);
(13) Hjorth et al. ( 2003)}

\end{table}

Even the great triumph of SN 2003dh has brought some new issues to
the fore.  The spectral evolution of SN 2003dh looks remarkably
like that of SN 1998bw.  How could that be since SN 2003dh was
associated with a classic gamma-ray burst and must have been
observed nearly down the jet axis and SN 1998bw was either
associated with an odd, very subluminous gamma-ray burst, or it
was seen substantially off axis.  The recent report of a
supernova-like spectrum in GRB 021211 by Della Valle et al (2003) also
adds a twist.  Here again, the supernova must be seen ``down the
pipe," but the velocities seem to be modest.  Clearly there is
still much to learn about the supernova gamma-ray  burst connection.

\section{Conclusions}

As we enter this workshop, we can point to several area of
critical interest.

Type Ia supernovae sometimes have significant polarization and
hence asymmetry.  This may yield clues to their binary origin.
Perhaps we are seeing evidence of how the combustion physics
proceeds. Perhaps there are hints, specifically, to the mechanism
of the crucial deflagration/detonation transition.

Much hard work has also shown that all core collapse explosions
are significantly polarized and hence asymmetric.  This means
that both the dynamics and the radiative processes (photons and
neutrinos!) are asymmetric.  An account of this asymmetry must
be made in the analysis of core collapse.

In particular, core collapse is an intrinsically shearing
environment, That makes it subject to the MRI, the resulting
turbulence, and hence to strong dynamo action and the exponential
growth of magnetic fields.  The implication is that rotation
and magnetic fields are intrinsic to the process of core
collapse for either neutron stars or black holes, for supernovae
or gamma-ray bursts.

Welcome to the brave new world of three-dimensional explosions!

\bigskip

{\it Acknowledgements:}  I am most grateful to Peter H\"oflich and
Pawan Kumar for hosting this wonderful meeting and putting in the
hard work before and after to make it scientifically successful
and stimulating.  I learned so much from so many colleagues over
the years, but would like to especially acknowledge those who
have recently led me into three-dimensional perspectives - Shizuka
Akiyama, Dietrich Baade, Vadim Gamezo, Andy Howell, Alexei Khokhlov, 
Itamar Lichtenstadt, Dave Meier, Elaine Oran, Peter Williams and
especially Lifan Wang and Peter H\"oflich. 
This work was supported in part by NSF AST-0098644 and by NASA NAG5-10766. 



\begin{thereferences}{99}

\bibitem{Akiyama} Akiyama, S. Wheeler, J. C., Meier, D. \& Lichtenstadt, I.
2003, \textit{Astrophysical Journa}l,
\textbf{584}, 954

\bibitem{1969Ap&SS...5..180A} Arnett, D.~W.\ 1969, \textit{Astrophysics \& Space Science}, 5,
180 

\bibitem{BlabusA}
Balbus, S. A. \& Hawley, J. F.  1991, \textit{Astrophysical Journal},
\textbf{376}, 214

\bibitem{BalbusB}
Balbus, S. A. \& Hawley, J. F. 1998, \textit{Review of Modern Physics},
\textbf{70}, 1

\bibitem{Blandford}
Blandford, R. D. \& Payne, D. G. 1982, \textit{Monthly Notices of the Royal
Astronomical Society}, \textbf{199}, 833

\bibitem{1997ApJ...483..675C} Clocchiatti, A.~et
al. 1997, \apj, \textbf{483}, 675

\bibitem{1996ApJ...459..547C} Clocchiatti, A., Wheeler, J.~C., Benetti,
S., \& Frueh, M.\ 1996, \apj, \textbf{459}, 547

\bibitem{2000ApJ...529..661C} Clocchiatti, A.~et
al.\ 2000, \apj, \textbf{529}, 661

\bibitem{Coburn}
Coburn, W. \& Boggs, S. E. 2003, \textit{Nature}, \textbf{423}, 415

\bibitem{2003A&A...406L..33D} Della Valle, M.~et
al.\ 2003, \aap, \textbf{406}, L33

\bibitem{duncan}
Duncan, R. C. \& Thompson, C. 1992, \apj, \textbf{392}, L9


\bibitem{1988AJ.....96.1941F} Filippenko, A.~V.\ 1988,
\aj, \textbf{96}, 1941

\bibitem{1990AJ....100.1575F}
Filippenko, A.~V., Porter, A.~C., \& Sargent, W.~L.~W.\ 1990, \aj,
\textbf{100},
1575

\bibitem{Frail}
Frail, D. A. et al. 2001, \apj, \textbf{562}, L55

\bibitem{1998Natur.395..670G} Galama, T.~J.~et al.\
1998, \nat, \textbf{395}, 670


\bibitem{2002MNRAS.332L..73G} Gal-Yam, A.,
Ofek, E.~O., \& Shemmer, O.\ 2002, \mnras, \textbf{332}, L73

\bibitem{2002AJ....124..417H} Hamuy, M.~et al.\ 2002,
\aj, \textbf{124}, 417

\bibitem{Hjorth}
Hjorth, J. et al. 2003, \nat, \textbf{423}, 847

\bibitem{Hoeflich:94D} H\"oflich, P.\ 1995, \apj, \textbf{443},
89 

\bibitem{Hoeflich99byIR}
H{\" o}flich, P., Gerardy, C.~L., Fesen, R.~A., \& Sakai, S.\ 2002, \apj,
\textbf{568}, 791

\bibitem{HofKho:1996} H\"oflich, P.~\&
Khokhlov, A.\ 1996, \apj, \textbf{457}, 500

\bibitem{hoef96}
H\"oflich, P., Khokhlov, A., Wheeler, J. C., Phillips, M. M. Suntzeff, N. B.
\& Hamuy, M.  1996, \apj, \textbf{472}, L81

\bibitem{hoef01}
H\"oflich, P., Khokhlov, A. \& Wang, L. 2001, in\textit{ Proc. of the 20th
Texas Symposium on Relativistic Astrophysics}, eds. J. C. Wheeler \&
H. Martel, (New York: AIP), 459

\bibitem{2002ApJ...568..779H} H{\" o}flich,
P.~\& Stein, J.\ 2002, \apj, \textbf{568}, 779

\bibitem{hoefwhe99}
H\"oflich, P., Wheeler, J. C., \& Wang, L. 1999, \apj, \textbf{521}, 179
 
\bibitem{howell}
Howell, D.~A., H{\" o}flich, P., Wang, L., \& Wheeler, J.~C.\ 2001, \apj,
\textbf{556}, 302

\bibitem{2003ApJ...593L..19K} Kawabata, K.~S.~et
al.\ 2003, \apjl,\textbf{ 593}, L19

\bibitem{Khokhlov(1991)} Khokhlov, A.~M.\ 1991, \aap, 
245, 114 

\bibitem{Khokhlov}
Khokhlov, A. \&  H\"oflich, P. 2001, in \textit{Explosive Phenomena
in Astrophysical Compact Objects}, eds. H.-Y, Chang, C.-H. Lee
\& M. Rho, AIP Conf. Proc. No. 556, (New York: AIP), p. 301

\bibitem{KhokhlovA}
Khokhlov, A. M. \& Oran, E. S. 1999, \textit{Combustion \& Flame},
\textbf{119}, 400

\bibitem{KhokhlovB}
Khokhlov, A. M., Oran, E. S. \& Thomas, G. O. 1999,
\textit{Combustion \& Flame}, \textbf{117}, 323

\bibitem{KhokhlovC}
Khokhlov  A.M., H\"oflich P. A., Oran E. S., Wheeler J. C.
Wang, L., \& Chtchelkanova, A. Yu. 1999, \apj, \textbf{524}, L107


\bibitem{kotake:2003} Kotake, K., Yamada, S. \& Sato, K. 2003, \apj, \textbf{595}, 304

\bibitem{Lentz:2001} Lentz, E.~J., Baron, E.,
Branch, D., \& Hauschildt, P.~H.\ 2001, \apj, \textbf{547}, 402

\bibitem{leonard}
Leonard, D.~C., Filippenko, A.~V., Barth, A.~J.,
\& Matheson, T.\ 2000, \apj, \textbf{536}, 239

\bibitem{leonardA}
Leonard, D.~C.~\& Filippenko, A.~V.\ 2001, \textit{Publications of the
Astronomical Society of the Pacific}, \textbf{113}, 920

\bibitem{leonardB}
Leonard, D.~C., Filippenko, A.~V., Ardila, D.~R.,
\& Brotherton, M.~S.\ 2001, \apj, \textbf{553}, 861

\bibitem{leonardC}
Leonard, D.~C., Filippenko, A.~V., Chornock, R. \& Foley, R. J.\ 2002,
\textit{PASP},
\textbf{114}, 1333 

\bibitem{Lyutikov}
Lyutikov, M., Pariev, V. I. \& Blandford, R. D. 2003, \apj, submitted
(astro-ph/0305410)

\bibitem{MacFayden}
MacFadyen, A., Woosley, S. E. \& Heger, A. 2001, \apj,\textbf{ 550}, 410

\bibitem{maeda}
Maeda, K., Nakamura, T., Nomoto, K., Mazzali, P., Patat, F.
\& Hachisu, I. 2002, \apj, \textbf{565}, 405

\bibitem{2003ApJ...591..316M} Marion, G.~H., H{\" o}flich, P., Vacca,
W.~D., \& Wheeler, J.~C.\ 2003, \apj, \textbf{591}, 316
 
\bibitem{2000ApJ...545..407M} Mazzali,
P.~A., Iwamoto, K., \& Nomoto, K.\ 2000, \apj, \textbf{545}, 407

\bibitem{meier}
Meier, D. L., Koide, S. \& Uchida, Y. 2001, \textit{Science}, \textbf{291},
84

\bibitem{1999ApJ...527..746M} Millard, J.~et al.\
1999, \apj, \textbf{527}, 746

\bibitem{Nugent:1997} Nugent, P., Baron, E.,
Branch, D., Fisher, A., \& Hauschildt, P.~H.\ 1997, \apj, \textbf{485}, 8

\bibitem{Ott} Ott, C. D., Burrows, A., Livne, E. \& Walder, R. 2003, \apj, in press
(astro-ph/0307472) 

\bibitem{pana}
Panaitescu. A. \& Kumar, P. 2001, \apj, \textbf{560}, L49

\bibitem{pk02}
Panaitescu, A. \& Kumar, P. 2002, \apj, \textbf{571}, 779

\bibitem{2001ApJ...555..900P} Patat, F.~et al.\ 2001,
\apj, \textbf{555}, 900

\bibitem{1996AJ....111..327R} Richmond, M.~W.~et
al.\ 1996, \aj, \textbf{111}, 327

\bibitem{stanek}
Stanek, K. Z. et al. 2003, \apj,\textbf{ 591}, L17

\bibitem{wang1}
Wang, L., Howell, D.~A., H{\" o}flich, P., \& Wheeler, J.~C.\ 2001, \apj,
\textbf{550}, 1030 

\bibitem{wang2}
Wang, L., Wheeler, J.~C., \& H\"oflich, P.\ 1997, \apjl, \textbf{476}, L27

\bibitem{wang3}
Wang, L., Wheeler, J.~C., Li, Z., \& Clocchiatti, A.\ 1996, \apj,
\textbf{467}, 435

\bibitem{wang4}
Wang, L. et al.\ 2002, \apj, \textbf{579}, 671 

\bibitem{wang5}
Wang, L. et al.\ 2003a, \apj, \textbf{591}, 1110   

\bibitem{wang6}
Wang, L. et al.\ 2003b, \apj,\textbf{ 592}, 457   

\bibitem{1996ssr..conf..241W} Wheeler,
J.~C.~\& Filippenko, A.~V.\ 1996, \textit{IAU Colloq.~145: Supernovae and
Supernova
Remnants}, 241

\bibitem{wheeler}
Wheeler, J.~C., Meier, D.~L. \& Wilson, J.~R.\ 2002, \apj, \textbf{568}, 807

\bibitem{wheeler2}
Wheeler, J. C., Yi, I., H\"oflich, P. \& Wang, L. 2000, \apj, \textbf{537},
810

\bibitem{zhang}
Zhang, W., Woosley, S. E. \& MacFadyen, A. I. 2003, \apj, \textbf{586}, 356


\end{thereferences}
\end{document}